\documentclass{PoS}

\usepackage{enumitem}
\usepackage{graphicx}
\usepackage{subcaption}

\usepackage{epsfig,times,color,cite,adjustbox,wrapfig,lipsum,booktabs,comment}
\usepackage{url}

\title{A Trigger Interface Board to manage trigger and timing signals in CTA Large-Sized Telescope and Medium-Sized Telescope cameras}

\ShortTitle{Trigger Interface Board for LST and MST in CTA}

\author{Pablo Pe\~nil$^{1}$, Luis \'{A}ngel Tejedor$^{1}$, Juan Abel Barrio$^{1}$, \speaker{Marcos L\'{o}pez}$^{1}$, for the CTA~consortium\\
{\footnotesize
\\$^{1}$Departamento de F\' {i}sica At\'{o}mica, Molecular y Nuclear, Universidad Complutense de Madrid, 28040 Madrid, Spain, }

E-mail: \email{ppenil@ucm.es}}

\abstract{One of the main goals of the Cherenkov Telescope Array (CTA) observatory is to improve the $\gamma$-ray detection sensitivity by an order of magnitude, compared to the current ground-based observatories. Widening the energy coverage down to 20 GeV and up to 300 TeV is also an important goal. This goal will be possible by using Large-Sized Telescopes (LSTs) for the energy range of 20--200 GeV, Medium-Sized Telescopes (MSTs) for 100 GeV--10 TeV, and Small-Sized Telescopes (SSTs) for energies above 5 TeV. The LSTs, which focus on the lowest energies, are operated in a region dominated by background events originated from the night sky background. To reduce such background events as much as possible, the LST cameras are only read out if at least two of them have been triggered in a short-time coincidence window. Such trigger is implemented for each LST camera in a dedicated module called Trigger Interface Board (TIB). In addition, the TIB is also used in MSTs equipped with the NectarCAM camera system to manage the different trigger and timing signals between LSTs and MSTs, as well as to monitor the different counting rates and dead-time of the cameras. It also assigns a time stamp to each event, which is recorded along with the information provided by the CTA global timing distribution system, based on the White Rabbit protocol. Therefore, the event arrival time can be determined in a redundant way. In this contribution, the main features and the technical performance of the TIB are presented.
}

\FullConference{35th International Cosmic Ray Conference -ICRC217-\\
		10-20 July, 2017\\
		Bexco, Busan, Korea}

\begin{document}

\section{Introduction}
The Cherenkov Telescope Array (CTA) will be the next ground-based $\gamma$-ray observatory, with the aim to improve both the sensitivity and the energy coverage with respect to current observatories. To achieve that aim, it is planned to build two different sets of telescopes in each hemisphere. The nature of the telescope is heterogeneous. Three different kinds of telescopes will be placed in the observatories according to their mirror diameters: Large-Sized Telescopes (LSTs), Medium-Sized Telescopes (MSTs) and Small-Sized Telescopes (SSTs).  
The LST camera \cite{uno_mazin} and NectarCAM \cite{2_nectarcam}, which is one of two MST camera designs, have been designed in a similar way, in terms of functionality and hardware implementation. However, they have a number of differences, since their designs are optimized for their corresponding target energy ranges. Despite their differences, both cameras must handle the trigger and timing signals they require essentially in the same manner. For that purpose, a Trigger Interface Board (TIB) has been deigned and produced. In the next sections, the TIB requirements and interfaces are presented. Additionally, a description of the main blocks of the TIB hardware/software implementation is shown, followed by a summarized description of its main functionalities. Finally, some results of the TIB performance are presented.

\section{TIB requirements}
All the telescopes in CTA are planned to work in coincidence. This involves some functional requirements that have to be implemented in all of them.
First, a common timing reference is required to recognize events observed by several telescopes corresponding to the same $\gamma$-ray shower. This timing reference is provided by a White Rabbit \cite{3_white} network. Every telescope receives a common 10-MHz clock signal and a pulse-per-second (PPS) signal at a dedicated White Rabbit node (hereinafter UCTS; Unified clock distribution Trigger time-Stamping), which is able to time-stamp events with sub-ns precision. 
In the case of the LSTs, with the aim of improving the sensitivity to the faint showers caused by low-energy $\gamma$-rays, the telescopes share their trigger information among neighbors in order to implement a hardware stereo trigger scheme. In this scheme, each LST camera only saves its data if an event has triggered that camera and at least one of the LST neighboring ones in a given coincidence time window. In this way, it is ensured that all the saved events are observed by at least two telescopes. This has the advantage of keeping the acquisition rate at affordable values (and hence the dead-time low), while maximizing the ratio of $\gamma$-ray-like events with respect to those induced by the night sky background (NSB). Therefore, a dedicated device is needed in each LST camera to coordinate the trigger signals coming from that camera and the ones from the neighboring telescopes. 

In addition to the $\gamma$-ray-like events detected by the trigger system, other causes must be able to trigger the camera readout process. For instance, one may want to read out the camera to perform flat-fielding calibrations, to take pedestals, or in general to check the correct functioning of the system. A system is required to centralize these trigger inputs from different origins and generate the final trigger signal, which effectively starts the camera readout process. 

All this set of functional requirements suggests the need of a device to centralize the trigger sources in the camera, implement the stereo scheme for LSTs and provide the trigger signal that will be time-stamped by the White Rabbit node in the camera with ns precision, for further array-level analysis. In this context, the TIB is the specific device that has been developed to accomplish these requirements in the LST camera and NectarCAM. Moreover, the TIB is the intelligent device that controls when the camera can be read out, taking into account the camera state machine and the busy state of the camera modules. 

As the camera-trigger centralizer, the TIB also provides a specific bit pattern, trigger type, for the data stream, which contains the trigger origin that caused the camera to be read out. In the case of a stereo event, the trigger type information is complemented with the set of neighbor telescopes that detected the event.

\section{TIB Interfaces}\label{interfaces}
\begin{figure}[ht]
\centering
\includegraphics[width=0.65\textwidth]{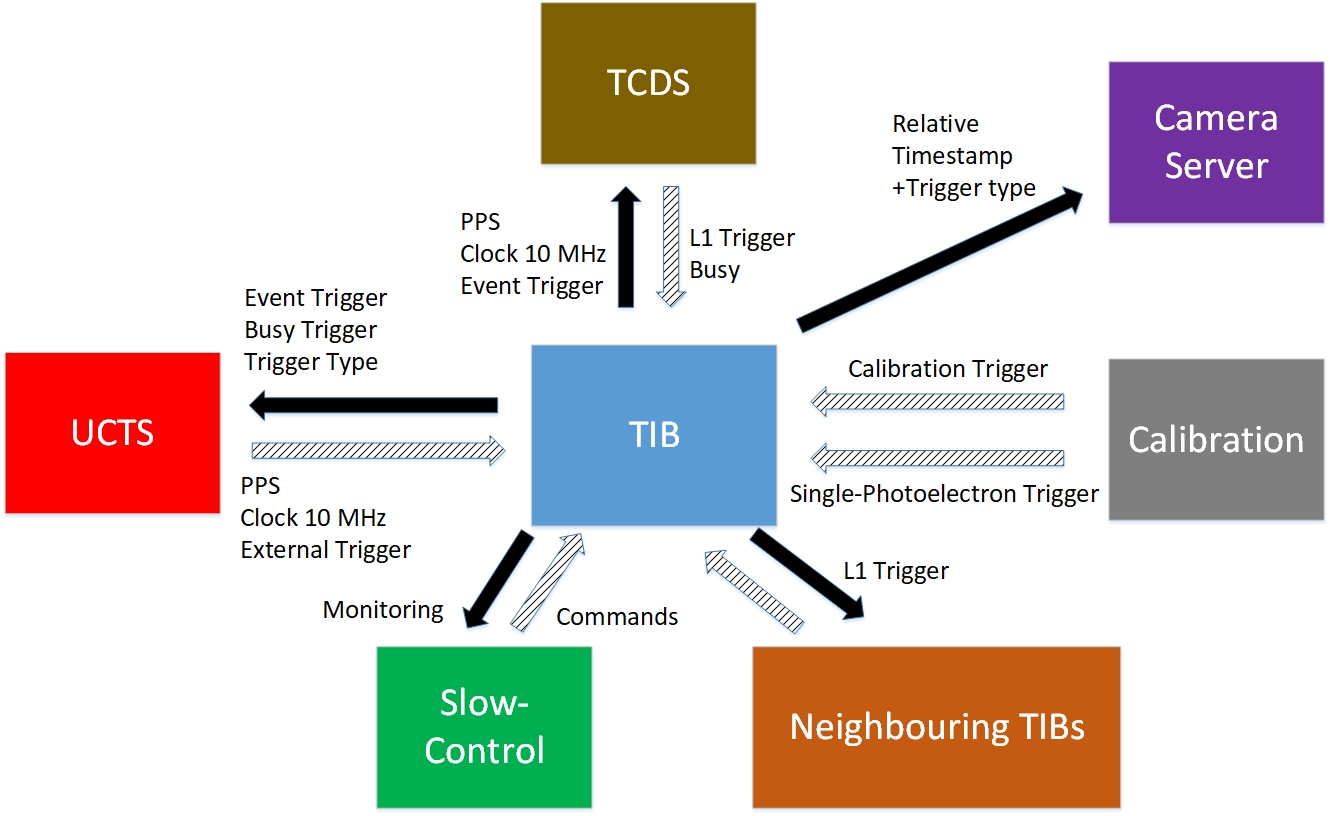}
\caption{\it  Scheme of the different subsystems connected with the TIB.}
\label{fig:tibinterfaces}
\end{figure}

For both LST camera and NectarCAM, the TIB interacts with a great variety of subsystems through a set of well-defined interfaces. Such structure is shown in Figure~\ref{fig:tibinterfaces}, where the different subsystems and the signals exchanged with the TIB are presented. Further details of the subsystems can be found in \cite{uno_mazin} and \cite{2_nectarcam}.

\subsection{UCTS}\label{ucts}
As is shown in Figure~\ref{fig:tibinterfaces}, the UCTS provides the PPS signal and 10-MHz clock signal to the TIB. These signals are distributed to the rest of the camera by the TIB and the trigger system (Section \ref{tcds}). In addition, the UCTS transmits an external trigger to the TIB, which can be used to start the camera readout process at a scheduled time.  

On the other hand, the TIB transmits the trigger signal that caused the camera read out to the UCTS, to get it time-stamped, and the trigger type associated to each trigger signal. After time-stamping the event, the UCTS sends it to the camera server, which gathers all the data and builds the event (Section \ref{cs}).   

\subsection{Trigger and Clock Distribution System}\label{tcds}
The LST camera and NectarCAM are modular cameras, each module being in charge of sampling and digitizing the information from seven pixels, also being able to decide on the camera read out, based on neighboring module information\footnote{The LST camera uses an analog trigger system \cite{4_analogo}, while NectarCAM can use either such analog trigger system or a digital trigger system \cite{5_digital}}. The trigger signal thus generated (L1) is transmitted to the TIB through a dedicated Trigger and Clock Distribution System (TCDS). These distribution systems are different depending on the kind of trigger used, specifically, analog trigger \cite{4_analogo} or digital trigger \cite{5_digital}.  

According to the camera state and the possible coincidences with neighboring LSTs, the TIB generates an event trigger, sending it back to the modules via the TCDS, so that the readout process can start.The TIB should not send any trigger to the modules while they are being read and busy. A busy signal is generated in the modules, keeping it in high state while the readout process is active. This busy signal is transmitted to the TIB through the TCDS. If a new L1 signal reaches the TIB while the busy is high, the TIB does not send the corresponding event trigger signal back to the modules. Finally, the TIB also resends the PPS and 10-MHz signals coming from the UCTS to the modules via the TCDS. 

\subsection{Camera Server}\label{cs}
The camera server is a computer which runs the event building process. A trigger bit pattern for each event is sent to the camera server from the TIB along with a reduced-precision (100 ns) time-stamp, that serves as cross-check of the high-precision UCTS time-stamp in the event building process. 

\subsection{Calibration}\label{cal}
Two different scenarios are considered for the calibration: Flat-fielding (FF) and single photoelectron (SPE) methods. Despite their different nature, they have common aspects: they must cause the readout process of the camera, they should not participate in the LST stereo trigger, and they must be clearly identified to be handled them properly in the camera server (for instance, these events should not be compressed). These special requirements are supported by the TIB, which receives the calibration trigger signals through dedicated lines, so they are handled in a special way. 

\subsection{Neighboring TIBs}\label{nt}
When the LSTs operate in the stereo trigger mode, the condition to decide if a $\gamma$-ray-like event has taken place is that, a configurable number of telescopes (two, two or more, three \ldots) have detected an image in their cameras and, consequently, they have fired their corresponding L1 in a time coincidence window. This scheme requires the L1 signals from several neighbor telescopes to take the final decision that will read out the cameras.

In the LSTs, this procedure is implemented as follows: in a given LST camera, the TIB processes the local L1 signal from the camera to which it belongs, and the L1 signals from the neighboring cameras. If at least one of the neighboring L1 signal arrives within typically 50 ns from the local L1, an event trigger is generated for that given LST camera and therefore their modules are read out. This procedure ensures that at least two LSTs are always read out whenever an event trigger is generated.

The TIBs in neighboring LST cameras are connected by means of optical fibers. Every time an LST camera receives a local L1 signal, it is broadcast to all the TIBs in their neighbors. Each TIB compensates for the latency in the arrival of the neighboring L1s, which changes with the pointing direction of the telescopes. This pointing direction is provided to each TIB by its slow-control interface  (Section \ref{sc}).

\subsection{Slow Control}\label{sc}
The TIB is a configurable device. In order to carry out some of its functionalities, a set of parameters need to be defined externally and transmitted to the TIB. For instance, the connection with the camera server is implemented by means of a TCP/IP socket, and the associated port and IP address are configured by the slow-control system. Other features related with the stereo trigger can be configured by the slow-control system, like the number of coincident LSTs or the width of the coincident window required to generate an event trigger. And, of course, several changes in the state machine are unchained by slow-control commands.

On the other hand, the slow-control of TIB not only enables to control the trigger behavior, but also provides useful information to the remote camera controller device. The TIB monitors the rates of different trigger inputs (local triggers generated in the camera, stereo triggers, calibration triggers and those occurring during the busy state known as busy triggers,...), the status of the connection with the camera server and  the temperature, and also provides alarm flags if a problem is detected.

All the slow-control of the TIB at the user level should be done by means of an OPC UA interface [6]. In the TIB microcomputer (Section \ref{architecture}), an OPC UA server is running, while in the remote camera controller device, an OPC UA client connects to this server, enabling the management of the functionalities previously described.  

\section{TIB Hardware Architecture}\label{architecture}
\begin{figure}[ht]
\centering
\includegraphics[width=0.5\textwidth]{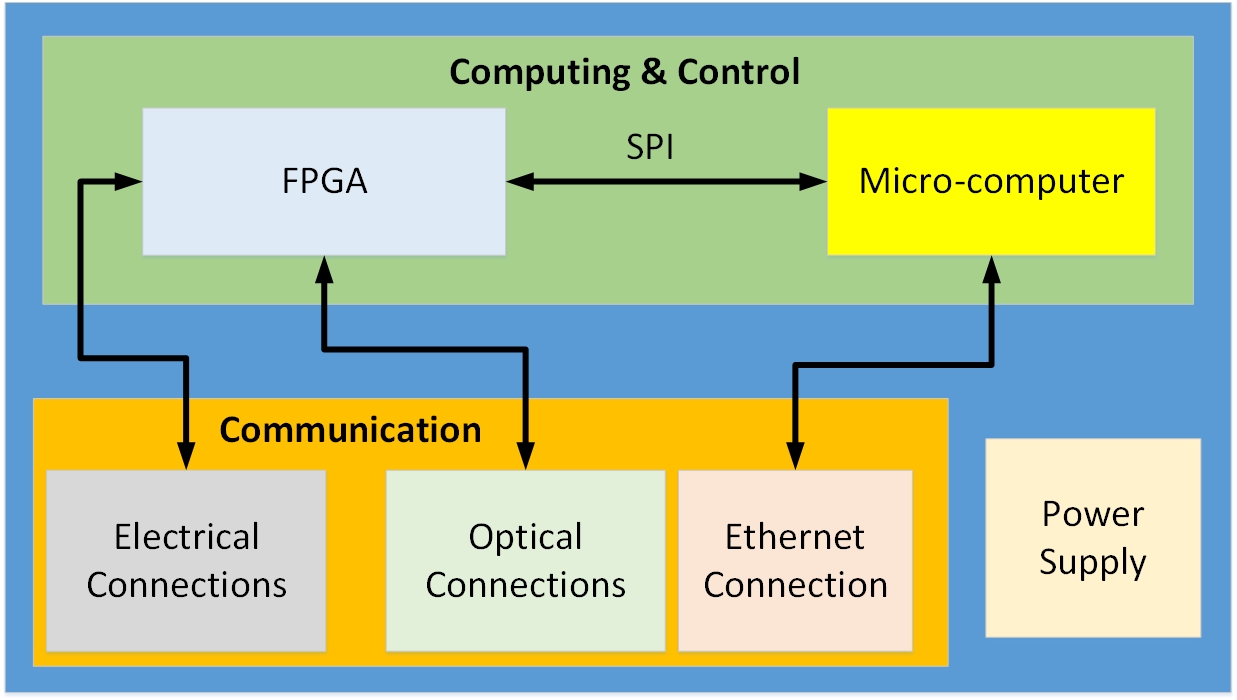}
\caption{\it  Scheme of the hardware architecture of the TIB.}
\label{fig:tibstructure}
\end{figure}

The hardware implementation of the TIB is composed of a set of blocks (Figure~\ref{fig:tibstructure}). These blocks can be mainly organized in two large groups. The first of them is related to the communications. Three different types of connections are used by the TIB. The electrical connections (RJ45 connectors using the LVDS standard) are used to receive/transmit the different signals between systems inside the camera, i.e. the UCTS, the TCDS and the SPE calibration box. A second type of connections is based on optical fibers. These connections are used to communicate with systems that are outside the camera, such as the FF calibration box and the TIBs in neighboring telescopes. Finally, the Ethernet connection handles the slow-control communication and is used to send the event trigger information to the camera server. 

All the functionalities of the TIB, apart from its interfaces, are actually implemented in the other two blocks: an FPGA which handles the fast signals, such as triggers, busy signals, etc (see Section \ref{functionality}) and builds the trigger type; and a microcomputer which is in charge of the slow-control system, the communication with the camera server, and the FPGA configuration at start-up. The FPGA and the microcomputer communicate with each other through a Serial Peripheral Interface (SPI) \cite{7_spi} link, with the microcomputer acting as a master.

\section{TIB Functional Behavior}\label{functionality}
\begin{figure}[ht]
\centering
\includegraphics[width=0.5\textwidth]{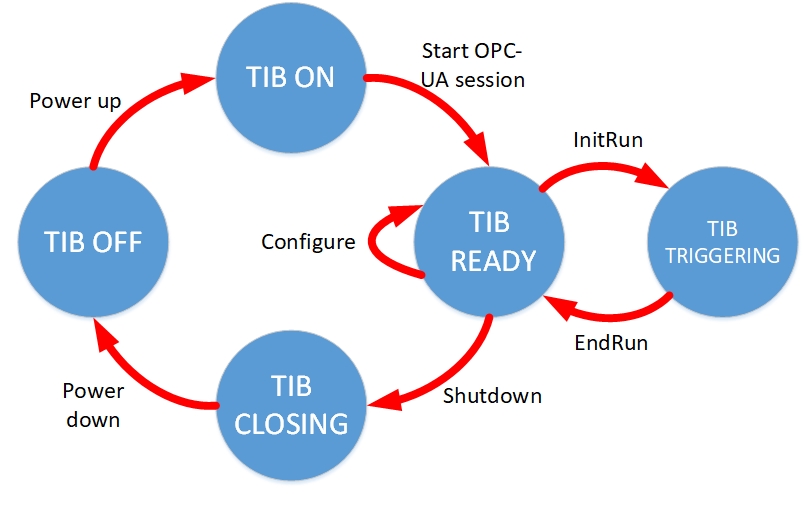}
\caption{\it  Simplified TIB state machine.}
\label{fig:tibmachine}
\end{figure}

The TIB behavior is modeled by the state machine shown in Figure~\ref{fig:tibmachine}. After powering on the TIB, two main actions are done in the `TIB ON` state. First, the microcomputer loads the firmware into the FPGA with the default parameters. Second, it starts the OPC UA server, which allows a client to control the behavior of the TIB and to monitor its main parameters. 

The FPGA firmware carries out the following functionalities:
\begin{itemize}[leftmargin=1cm]  
\item Implementing the management of all the trigger inputs, enabling or disabling the connections (calibration, busy \ldots), according to the camera state machine and other configurations.   
\item Enabling/disabling the PPS distribution to the TCDS, depending on the camera state. 
\item Controlling the parameters of the stereo scheme, like the coincident window or the minimum number of telescopes in coincidence. 
\item Configuring the asynchronous delay of the event trigger signal with respect to the local L1 trigger.
\item Managing the rates of all trigger inputs and outputs.
\item Building the trigger type information associated to the event and sending it to the UCTS.
\item Create its own event information by combining the trigger type and a 100-ns precision time-stamp. This information is read out by the microcomputer, which sends it to the camera server.
\end{itemize}

Once all the parameters are configured properly, the TIB is ready to start triggering the camera by running the following procedure: when every camera module is ready, the operator sends a slow-control command to the TIB, indicating that the next PPS signal from the UCTS must be the first one to be distributed. The leading edge of this PPS will be considered as the time origin at the TIB and at all the camera modules and, since that moment, the modules are ready to accept triggers. 

While the TIB is processing triggers, the microcomputer sends the event information generated by the TIB (trigger type + time-stamp) to the camera server. 

\begin{figure}[ht]
\centering
\begin{subfigure}[b]{0.4\textwidth}
        \includegraphics[width=\textwidth]{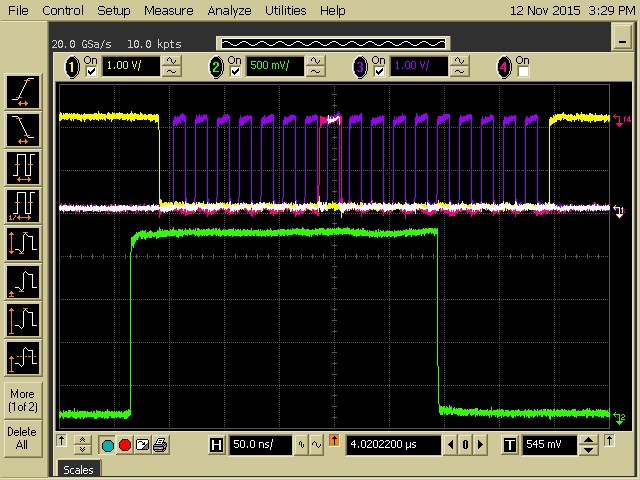}
        \caption{SPI communication}
        \label{fig:spi}
    \end{subfigure}
    \begin{subfigure}[b]{0.4\textwidth}
        \includegraphics[width=\textwidth]{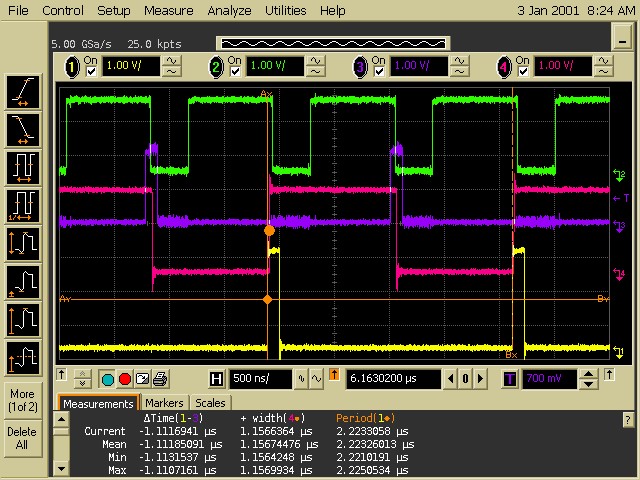}
        \caption{Trigger Type}
        \label{fig:triggertype}
    \end{subfigure}
    \caption{(\emph{a}) Event trigger (green) and SPI trigger type transmission signals: chip select that enables SPI transmission (yellow), clock (purple), and data (magenta). (\emph{b}) Event trigger (yellow), busy signal (magenta), busy triggers (purple) and chip select  corresponding to the trigger type transmission (green).}\label{fig:results}
\end{figure}

\section{Results}
As the TIB is a subsystem interfacing many others in the camera, most of its tests consist of verification of its functionalities at integration tests with other subsystems. Currently, there are TIBs working at different LST camera and NectarCAM demonstration test benches at different institutes in several countries. These test benches include a UCTS, several camera modules, a TCDS, etc, depending on the test purposes. Some of these tests have been carried out with a camera server emulator, in order to verify the coherence between the trigger type, the time stamp generated by the TIB and the event information produced by the UCTS. The results obtained so far have proven such coherence. The maximum trigger processing rate reached by the TIB is around 16 kHz. This value is mainly constrained by the data delivery to the camera server, handled by the microcomputer. Otherwise, the trigger processing rates could reach values of several MHz, depending on the busy state duration. 

Apart from the results obtained at the integration tests, some of the main features achieved are:
\begin{itemize}[leftmargin=1cm] 
\item Possibility to asynchronously adjust the delay between the L1 trigger input and event trigger output, with a 3850-ns adjustable range and around 1-ns jitter in the largest delay case. 
\item Trigger type generation and delivery to the UCTS through a SPI link in less than 450 ns (Figure~\ref{fig:spi}).
\item Busy handling extending the busy time if a trigger takes place during the busy transition from high to low, to avoid cutting the event trigger pulse (Figure~\ref{fig:triggertype}).
\item 10-MHz and PPS delivery from the UCTS to the TCDS, adding ~10-ps and ~30-ps jitter, respectively.
\end{itemize}

\section{Conclusion}
A Trigger Interface Board has been designed in order to equip the camera of the LSTs and the NectarCAM of MSTs. Its purpose is to handle the trigger and timing signals required in both cameras. It will also implement the stereo trigger scheme among LSTs.  Several prototypes have been constructed and successfully characterized, so it is finally ready to be installed in the preproduction version of the LSTs and the NectarCAM-MSTs. 

\section{Acknowledgements}
This work was conducted in the context of the LST project and the NectarCAM work package, whose members provided us with a lot of help. In this sense, we specially want to thank the CTA analog trigger team. We gratefully acknowledge financial support from the agencies and organizations listed here :
\url{http://www.cta-observatory.org/consortium_acknowledgments}

\end{document}